\documentclass[twocolumn,english,prl,floatfix,citeautoscript,nofootinbib]{revtex4}
\usepackage{amsfonts}
\usepackage{amsbsy}
\usepackage{latexsym,epsfig,graphicx}
\usepackage{dcolumn}
\usepackage{subfigure}
\usepackage{comment}
\usepackage{color}
\usepackage[colorlinks,urlcolor=blue,citecolor=blue]{hyperref}
\usepackage{amstext}
\usepackage{amssymb}
\usepackage{setspace}

\usepackage{amsmath}

\usepackage{bm}

\begin{document}

\title{$\mathcal{PT}$-Symmetry Enhanced Berezinskii-Kosterlitz-Thouless
Superfluidity}
\author{Yun-Mei Li}
\affiliation{Department of Physics, The University of Texas at Dallas, Richardson, Texas
75080, USA}
\author{Xi-Wang Luo}
\affiliation{Department of Physics, The University of Texas at Dallas, Richardson, Texas
75080, USA}
\author{Chuanwei Zhang}
\email{chuanwei.zhang@utdallas.edu}
\affiliation{Department of Physics, The University of Texas at Dallas, Richardson, Texas
75080, USA}

\begin{abstract}
Berezinskii-Kosterlitz-Thouless (BKT) transition, the topological phase
transition to a quasi-long range order in a two-dimensional (2D) system, is
a hallmark of low-dimensional topological physics. The recent emergence of
non-Hermitian physics, particularly parity-time ($\mathcal{PT}$) symmetry,
raises a natural question about the fate of low-dimensional orders (e.g.,
BKT transition) in the presence of complex energy spectrum. Here we
investigate the BKT phase transition in a 2D degenerate Fermi gas with an
imaginary Zeeman field obeying $\mathcal{PT}$-symmetry. Despite complex
energy spectrum, $\mathcal{PT}$-symmetry guarantees that the superfluid
density and many other quantities are real. Surprisingly, the imaginary
Zeeman field enhances the superfluid density, yielding higher BKT transition
temperature than that in Hermitian systems. In the weak interaction region,
the transition temperature can be much larger than that in the strong
interaction limit. Our work showcases a surprising interplay between
low-dimensional topological defects and non-Hermitian effects, paving the
way for studying non-Hermitian low-dimensional phase transitions.
\end{abstract}

\maketitle

Berezinskii-Kosterlitz-Thouless (BKT) transition, first discovered in the
two-dimensional (2D) XY spin model~\cite%
{VLBerezinskii1,VLBerezinskii2,Krosterliz,Thouless}, is a cornerstone for
studying low-dimensional condensed matter physics~\cite{JMKosterlitz}. In 2D
systems, while Mermin--Wagner theorem forbids the emergence of a true long
range order due to thermal fluctuations, BKT physics permits a topological
phase transition to a quasi-long range order of topological defects (i.e.,
vortices) at very low temperature. Specifically, across a critical
temperature $T_{\mathrm{BKT}}$, free vortices bind together spontaneously
and form bound vortex-antivortex (V-AV) pairs, giving rising to a quasi-long
range order. BKT transitions have been experimentally studied in a wide
range of systems, including liquid helium films \cite{Bishop}, 2D magnets
\cite{MHeinrich,UTutsch,ZHu,HLi}, superconducting thin films \cite%
{AFHebard,KEpstein,DJResnick,MMondal}, and 2D atomic hydrogen \cite{Safonov}%
, where only macroscopic properties of the systems can be measured. In
recent years, ultracold atomic superfluids~have emerged as a versatile
playground for the studies of BKT physics \cite{ZHadzibabic, Desbuquois,
Clade, Tung, Murthy, Christodoulou, Sobirey,
Mitra,SSBotelho,Bertaina,MGong2,JPADevreese,YXu,Wu2015} with access to the
underlying microscopic phenomena such as the visualization of the
proliferation of free vortices. Experimental signatures of BKT physics have
been observed in harmonically trapped Bose and Fermi gases \cite%
{ZHadzibabic, Desbuquois, Clade, Tung, Murthy, Christodoulou, Sobirey, Mitra}%
, and\ the recent realization of ultracold superfluids in box potentials
\cite{Chuu, Gaunt, Mukherjee, Hueck, Patel} offers a uniform platform for
exploring BKT transitions.

So far the BKT transition has only been investigated in isolated systems
governed by Hermitian Hamiltonians. The recent emergence of non-Hermitian
physics, in particular parity-time ($\mathcal{PT}$) symmetry \cite%
{CMBender,CMBender2}, in photonic and ultracold atomic systems \cite%
{RGanainy, Feng, MMuller, PPeng, JLi, SLapp, Takasu} raises a natural
question on the fate of BKT transition in open systems, where the coupling
with external environments leads to gain or loss that are generally
described by non-Hermitian Hamiltonians. While the energy eigenvalues for
non-Hermitian Hamiltonians are complex in general, the $\mathcal{PT}$%
-symmetry of a Hamiltonian ensures either all real ($\mathcal{PT}$%
-symmetric) or complex conjugate pairs ($\mathcal{PT}$-broken) of complex
eigenvalues, with two regions separated by an exceptional point \cite%
{CMBender,CMBender2}. In ultracold atomic gases, the physical realization
\cite{PPeng, JLi, SLapp, Takasu} as well as associated single particle
physics \cite{YAshida, XWLuo2, TYoshida, YXu2, LLi, JXu, TLiu} of $\mathcal{%
PT}$-symmetry have been investigated in both theory and experiment. Recently
non-Hermitian fermionic superfluidity at zero temperature with
complex-valued interactions or $\mathcal{PT}$-symmetric pairing states have
been studied in theory \cite{NMChtchelkatchev,AGhatak,KYamamoto,LZhou}.
However, non-Hermitian effects on the finite temperature BKT transitions in
low-dimensional quantum systems remain unexplored.

In this Letter, we study the BKT phase transition in a 2D attractive Fermi
gas in the presence of real in-plane and imaginary out-of-plane Zeeman
fields that obey $\mathcal{PT}$-symmetry. In the $\mathcal{PT}$-symmetric
region, quasiparticle eigenenergies are real and the BKT physics is similar
as the Hermitian system. Interestingly, despite complex quasiparticle
eigenenergies in the $\mathcal{PT}$-broken region, many physical quantities,
such as order parameter and superfluid density are still real and oscillate
with the imaginary Zeeman field. The superfluid density in the $\mathcal{PT}$%
-broken region is larger than that in the $\mathcal{PT}$-symmetric region,
although the order parameter only changes slightly. As a result, the BKT
temperature is higher in the $\mathcal{PT}$-broken region. In the weak
coupling regime, the BKT transition temperature can be larger than the
largest value $E_{F}/8$ for Hermitian system that is achieved in the strong
coupling limit. The BKT transition temperature also shows an oscillating
behavior with respect to the imaginary Zeeman field. Such surprising
enhancement of BKT transition temperature showcases the interplay between
non-Hermiticity and low-dimensional orders and phase transitions, which may
open an avenue for exploring low-dimensional non-Hermitian physics.

\emph{Effective action: }Consider a 2D degenerate Fermi gas with two atomic
hyperfine states denoted as the pseudospin, where 2D confinement can be
implemented by a strong harmonic trap or a deep optical lattice along the
third dimension. The Fermi gas is subjected to real in-plane ($h_{x}$) and
imaginary out-of-plane ($ih_{z}$) Zeeman fields. The many-body Hamiltonian
reads%
\begin{eqnarray}
H &=&\int d\mathbf{r}\hat{\Psi}^{\dagger }(\mathbf{r})H_{s}(\mathbf{p})\hat{%
\Psi}(\mathbf{r})  \nonumber \\
&&-U\int d\mathbf{r}\hat{\Psi}_{\uparrow }^{\dagger }(\mathbf{r})\hat{\Psi}%
_{\downarrow }^{\dagger }(\mathbf{r})\hat{\Psi}_{\downarrow }(\mathbf{r})%
\hat{\Psi}_{\uparrow }(\mathbf{r}),  \label{eq1}
\end{eqnarray}%
where the single-particle Hamiltonian $H_{s}(\mathbf{p})=\mathbf{p}%
^{2}/2m-\mu +H_{z}$ with the chemical potential $\mu $, Zeeman field $%
H_{z}=h_{x}\sigma _{x}+ih_{z}\sigma _{z}$ ($h_{x},h_{z}>0$), and Pauli
matrices $\mathbf{\sigma }$. $H_{s}(\mathbf{p})$ preserves a $\mathcal{PT}$%
-symmetry: $[H_{s},\mathcal{PT}]=0$ with $\mathcal{P}=\sigma _{x}$ and $%
\mathcal{T}=\mathcal{K}$, where $\mathcal{K}$ denotes the complex conjugate.
The exceptional point between $\mathcal{PT}$-symmetric and -broken regions
is at $h_{x}=h_{z}$. $U$ is the attractive interaction strength. $\hat{\Psi}(%
\mathbf{r})=[\hat{\Psi}_{\uparrow }(\mathbf{r}),\hat{\Psi}_{\downarrow }(%
\mathbf{r})]^{T}$ and $\hat{\Psi}_{\nu }^{\dagger }(\mathbf{r})$ ($\hat{\Psi}%
_{\nu }(\mathbf{r})$) creates (annihilates) a fermionic atom at $\mathbf{r}$.

In quantum field theory, the partition function at temperature $T=1/\beta $
can be written as a path integral $Z=\int D\bar{\phi}D\phi e^{-S_{\mathrm{eff%
}}[\bar{\phi},\phi ]}$ with the effective action $S_{\mathrm{eff}%
}=\int_{0}^{\beta }d\tau \int d\mathbf{r}(\frac{|\phi |^{2}}{U}+\sum_{%
\mathbf{p}}\xi _{\mathbf{p}})-\frac{1}{2}\mathrm{Tr}[\ln G^{-1}]$, and $\xi
_{\mathbf{p}}=\mathbf{p}^{2}/2m-\mu $. Here $\phi (\mathbf{r},\tau )$ is the
superfluid order parameter and the trace $\mathrm{Tr}$ is over momentum $%
\mathbf{p}$, imaginary time $\tau $, and the Nambu index. The inverse Green
function $G^{-1}=-\partial _{\tau }-H_{\mathrm{BdG}}$, with the
Bogoliubov-de Gennes (BdG) Hamiltonian in the Nambu basis given by
\begin{equation}
H_{\mathrm{BdG}}=\left(
\begin{array}{cccc}
\xi _{\mathbf{p}}+ih_{z} & h_{x} & 0 & -\phi  \\
h_{x} & \xi _{\mathbf{p}}-ih_{z} & \phi  & 0 \\
0 & \phi ^{\ast } & -\xi _{\mathbf{p}}-ih_{z} & -h_{x} \\
-\phi ^{\ast } & 0 & -h_{x} & -\xi _{\mathbf{p}}+ih_{z}%
\end{array}%
\right) .  \label{eq2}
\end{equation}

It is known that a large Zeeman field can induce finite-momentum Cooper
pairing, \textit{i.e.}, Fulde--Ferrell--Larkin--Ovchinnikov (FFLO) states
\cite{YLiao,Ong}. Here we are interested in the region with a small $h_{x}$.
Assuming $\phi (\mathbf{r},\tau )=\Delta e^{i\mathbf{q}\cdot \mathbf{r}}$
and using the mean-field saddle point $\frac{\partial S_{\mathrm{eff}}}{%
\partial \Delta }=0$, $\frac{\partial S_{\mathrm{eff}}}{\partial \mathbf{q}}%
=0$, we find that $\mathbf{q}=0$ for a large imaginary field $ih_{z}$.
Therefore we only consider zero momentum BCS pairing in this paper. The
interaction parameter is regularized as $1/U=\sum_{\mathbf{k}}1/(\hbar
^{2}k^{2}/m+E_{B})$, where $E_{B}$ is the two-body binding energy that can
be varied by tuning the \textit{s}-wave scattering length using Feshbach
resonance or the barrier height along the $z$ direction.

\begin{figure}[tbp]
\centering
\includegraphics[width=0.48\textwidth]{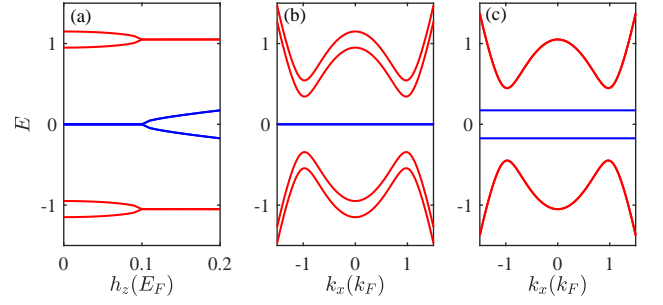}\newline
\caption{(a) BdG quasiparticle energy spectrum with respect to $h_{z}$ at $%
k_{x}=k_{y}=0$. (b, c): Quasiparticle spectrum at $h_{z}=0$ (b) and $%
h_{z}=0.2E_{F}$ (c) along the $k_{x}$ axis. $h_{x}=0.1E_{F}$. $\Delta $ and $%
\protect\mu $ are obtained self-consistently. The red and blue lines
correspond to real and imaginary parts of the quasiparticle energies.}
\label{fig1}
\end{figure}

The quasiparticle spectrum has four branches $E=\pm E_{\mathbf{p}}\pm \sqrt{%
h_{x}^{2}-h_{z}^{2}}$ with $E_{\mathbf{p}}=\sqrt{|\phi |^{2}+\xi _{\mathbf{p}%
}^{2}}$, which are real (complex conjugate pairs) in the $\mathcal{PT}$%
-symmetric (-broken) region $h_{z}<h_{x}$ ($h_{z}>h_{x}$), as shown in Fig.~%
\ref{fig1}. This property of the quasiparticle spectrum is guaranteed by the
$\mathcal{PT}$-symmetry, which is still preserved for the BdG Hamiltonian
with $\mathcal{P}=\sigma _{x}\tau _{0}$ and $\mathcal{T}=\mathcal{K}$. Here $%
\boldsymbol{\tau }$ is the Pauli matrix on the particle-hole space. The BdG
Hamiltonian also possesses a particle-hole symmetry $\mathcal{C}H_{\mathrm{%
BdG}}(\mathbf{p})\mathcal{C}^{-1}=-H_{\mathrm{BdG}}^{\dagger }(-\mathbf{p})$
with $\mathcal{C}=\sigma _{x}\tau _{y}$ and $\mathcal{C}^{2}=1$. Due to the
symmetry constraints, the effective action at the saddle point
\begin{equation}
S_{0}=\frac{\beta \Delta ^{2}}{U}+\sum_{\mathbf{p}}[\beta \xi _{\mathbf{p}%
}-\ln (2\cosh \beta h_{\mathrm{eff}}+2\cosh \beta E_{\mathbf{p}})],
\label{eq3}
\end{equation}%
is always real even in the $\mathcal{PT}$-broken region, where $h_{\mathrm{%
eff}}=\sqrt{h_{x}^{2}-h_{z}^{2}}$. Similar result also applies to other
quantities such as the partition function.

To study the phase fluctuations, we set $\phi (\mathbf{r},\tau )=\Delta
e^{i\theta (\mathbf{r},\tau )}$, where $\theta (\mathbf{r},\tau )$ is the
phase fluctuation around the saddle point. Around the saddle point, the
effective action can be written as $S_{\text{eff}}=S_{0}+S_{\text{fluc}}$
with%
\begin{equation}
S_{\text{fluc}}=\frac{1}{2}\int d^{2}\mathbf{r}[J(\nabla \theta
)^{2}+P(\partial _{\tau }\theta )^{2}-Q(i\partial _{\tau }\theta )],
\label{eq4}
\end{equation}%
where
\begin{equation}
J=\frac{1}{4m}\sum_{\mathbf{p}}[n_{\mathbf{p}}-\frac{\beta \mathbf{p}^{2}}{8m%
}\sum_{i=\pm }\mathrm{sech}^{2}\frac{\beta E_{i}}{2}]  \label{SupDen}
\end{equation}%
is the superfluid density, $P=\frac{1}{8}\sum_{\mathbf{p}}[\frac{\Delta ^{2}%
}{E_{\mathbf{p}}^{3}}\sum_{i=\pm }\tanh \frac{\beta E_{i}}{2}+\frac{\beta
\xi _{\mathbf{p}}^{2}}{2E_{\mathbf{p}}^{2}}\sum_{i=\pm }\mathrm{sech}^{2}%
\frac{\beta E_{i}}{2}]$ is the compressibility, $Q=\sum_{\mathbf{p}}n_{%
\mathbf{p}}$, $n_{\mathbf{p}}=1-\frac{\xi _{\mathbf{p}}}{2E_{\mathbf{p}}}%
\sum_{i=\pm }\tanh \frac{\beta E_{i}}{2}$, and $E_{\pm }=E_{\mathbf{p}}\pm
h_{\mathrm{eff}}$. In the $\mathcal{PT}$-broken region, the terms $%
\sum_{i=\pm }\tanh \frac{\beta E_{i}}{2}=\frac{2\sinh (\beta E_{\mathbf{p}})%
}{\cosh (\beta E_{\mathbf{p}})+\cos (\beta h^{\prime })}$ in $n_{\mathbf{p}}$
and $\sum_{i=\pm }\mathrm{sech}^{2}\frac{\beta E_{i}}{2}=\frac{4\left[
1+\cosh (\beta E_{\mathbf{p}})\cos (\beta h^{\prime })\right] }{1+2\cosh
(\beta E_{\mathbf{p}})\cos (\beta h^{\prime })+[\cosh (2\beta E_{\mathbf{p}%
})+\cos (2\beta h^{\prime })]/2}$ \cite{SM} with real $h^{\prime }=\sqrt{%
h_{z}^{2}-h_{x}^{2}}$, therefore $J$ is purely real as guaranteed by the $%
\mathcal{PT}$ symmetry. In fact, the parameters $J$, $P$, $Q$ are all real
and positive, showing the superfluid phase is stable even in the $\mathcal{PT%
}$-broken region.

We can decompose the phase $\theta $ into a static vortex part $\theta _{v}(%
\mathbf{r})$ and a time-varying spin wave part $\theta _{sw}(\mathbf{r},\tau
)$, leading to $S_{\text{fluc}}=S_{v}+S_{sw}$ in the action of the phase
fluctuation. Here $S_{v}=(1/2)\int d^{2}\mathbf{r}J(\nabla \theta _{v})^{2}$%
, $S_{sw}=(1/2)\int d^{2}\mathbf{r}[J(\nabla \theta _{sw})^{2}+P(\partial
_{\tau }\theta _{sw})^{2}-Q(i\partial _{\tau }\theta _{sw})]=\sum_{\mathbf{k}%
}\ln [1-\exp (-\beta \omega _{\mathbf{k}})]$, $\omega _{\mathbf{k}}=c|%
\mathbf{k}|$, and $c=\sqrt{J/P}$ is the sound speed. With the phase
fluctuation, the parameters $\Delta $ and $\mu $ can be calculated
self-consistently by solving the equations:
\begin{equation}
\frac{\partial S_{0}}{\partial \Delta }=0,\text{ }\frac{\partial \Omega }{%
\partial \mu }=-n  \label{TD}
\end{equation}%
with $\Omega =\left( S_{0}+S_{sw}\right) /\beta $ the thermodynamic
potential. The atom density $n=mE_{F}/\pi $. $E_{F}=\hbar ^{2}k_{F}^{2}/2m$
and $k_{F}^{-1}$ are chosen as the energy and length units, respectively.
The mean-field transition temperature $T_{\mathrm{MF}}$ can be obtained from
the saddle point equations with $\Delta =0$, below which free vortices
emerge ($\Delta \neq 0$) without phase coherence. The BKT phase transition
temperature is determined self-consistently by \cite%
{VLBerezinskii1,VLBerezinskii2,Krosterliz,Thouless}
\begin{equation}
T_{\text{\textrm{BKT}}}=\frac{\pi }{2}J(\Delta ,\mu ,T_{\text{\textrm{BKT}}%
}).  \label{eq5}
\end{equation}%
Across $T_{\text{\textrm{BKT}}}$, phase incoherent free vortices and
anti-vortices transit to phase-coherent bound V-AV pairs, yielding a
superfluid with a quasi-long range order. When the temperature further
decreases below another critical temperature $T_{\text{vortex}}=0.3J(\Delta
,\mu ,T_{\text{vortex}})$, V-AV pairs form a square lattice~\cite%
{BIHalperin,DRNelson,APYoung}.

\begin{figure}[t]
\centering
\includegraphics[width=0.5\textwidth]{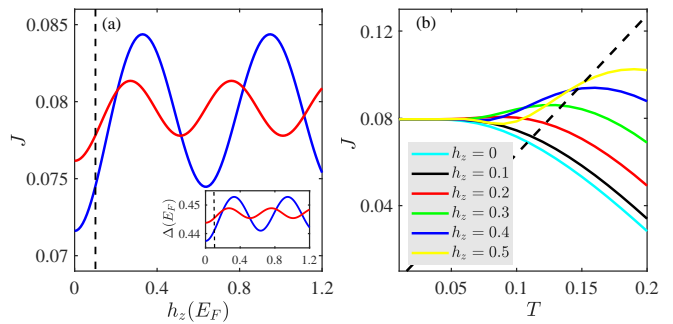}\newline
\caption{The dependence of superfluid density $J$ on $h_{z}$ (a) and $T$
(b). $E_{b}=0.1E_{F}$, $h_{x}=0.1E_{F}$. In (a), the blue and red lines
denote the temperature $T=0.1$ and $0.08$, respectively. The inset shows the
order parameter $\Delta $ with respect to $h_{z}$. In (b), the intersection
point between the black dashed line $y=2T/\protect\pi $ and superfluid
density line corresponds to the BKT transition point.}
\label{fig2}
\end{figure}

\emph{BKT transition with }$\mathcal{PT}$\emph{\ symmetry}: In Fig.~\ref%
{fig2} (a), we plot the superfluid density $J$ with respect to $h_{z}$ at a
given $h_{x}=0.1E_{F}$. In the $\mathcal{PT}$-symmetric region $h_{z}<h_{x}$%
, the single-particle Hamiltonian can be transformed to a Hermitian
Hamiltonian $\hat{S}^{-1}(\xi _{\mathbf{p}}+h_{x}\sigma _{x}+ih_{z}\sigma
_{z})\hat{S}=\xi _{\mathbf{p}}+h_{\mathrm{eff}}\sigma _{x}$, with $\hat{S}%
=e^{\alpha \sigma _{y}}$ and $\alpha =\frac{1}{4}\ln \frac{h_{x}-h_{z}}{%
h_{x}+h_{z}}$. Similarly, the transformation operator $\hat{S}^{\prime
}=e^{\alpha \sigma _{y}\tau _{0}}$ has the same effect on the BdG
Hamiltonian (\ref{eq2}). It is known that the increase (decrease) of Zeeman
field reduces (enhances) the order parameter in Hermitian systems,
accompanied by the decrease (increase) of the superfluid density and the BKT
transition temperature. Fig.~\ref{fig2} (a) shows that in the $\mathcal{PT}$%
-symmetric region the superfluid density increases as $h_{z}$ increases,
because the effective Zeeman field $h_{\mathrm{eff}}$ decreases. The 2D
Fermi gas in the $\mathcal{PT}$-symmetric region behaves the same as a
Hermitian system with a real Zeeman field.

In the $\mathcal{PT}$-broken region $h_{z}>h_{x}$, the quasiparticle
spectrum is complex, but forms complex conjugate pairs due to the $\mathcal{%
PT}$ symmetry (Fig.~\ref{fig1}), which lead to real effective action and
superfluid density. The superfluid density first increases, then oscillates
with respect to $h_{z}$ (Fig.~\ref{fig2} (a)) at fixed temperatures, which
do not occur in BKT transitions for Hermitian systems. Such oscillation
originates from the pair of complex conjugate quasiparticle eigenvalues,
which lead to a periodic oscillation of the effective action at saddle point
\begin{equation}
S_{0}=\frac{\beta \Delta ^{2}}{U}+\sum_{\mathbf{p}}[\beta \xi _{\mathbf{p}%
}-\ln (2\cos \beta h^{\prime }+2\cosh \beta E_{\mathbf{p}})]
\end{equation}%
with the periodicity $\delta h^{\prime }=2\pi T$. Other parameters $J$, $P$,
$Q$, $\Delta $ (inset in Fig.~\ref{fig2}(a)), $\mu $ show a similar periodic
dependence, as we see from the expression of $J$ in the $\mathcal{PT}$%
-broken region (the paragraph below Eq. (\ref{SupDen})). The oscillation
amplitudes are smaller for lower temperatures, and the superfluid density in
the $\mathcal{PT}$-broken region is larger than that in the $\mathcal{PT}$%
-symmetric region. The dependence of the superfluid density on the
temperature is plotted in Fig.~\ref{fig2}(b) for different $h_{z}$. At high
temperature the superfluid density in the $\mathcal{PT}$-symmetric (-broken)
region is smaller (larger) than that at the exceptional point $h_{z}=h_{x}$,
which corresponds to the superfluid without Zeeman field.

\begin{figure}[t]
\centering
\includegraphics[width=0.5\textwidth]{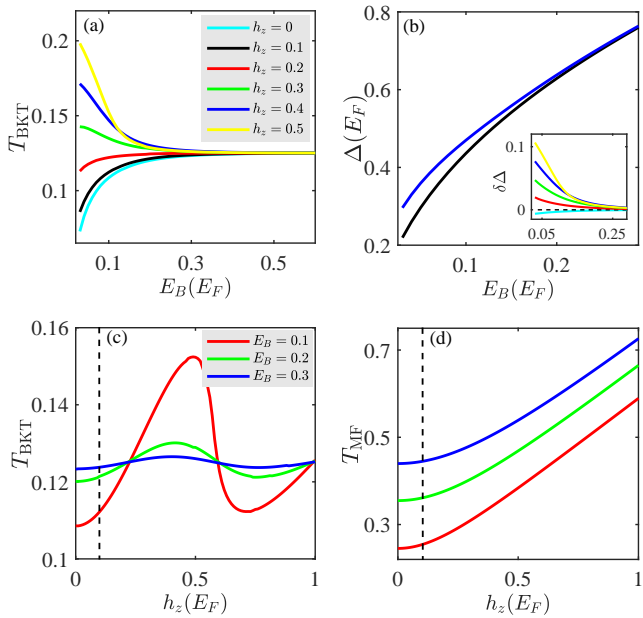}\newline
\caption{(a) Dependence of $T_{\mathrm{BKT}}$ on the binding energy $E_{B}$.
(b) Dependence of the order parameter $\Delta $ at $T_{\mathrm{BKT}}$ on $%
E_{B}$. The inset shows the $\Delta $ difference between $h_{z}=0.1E_{F}$
and other $h_{z}$ values. The color notation is the same as (a). (c,d)
Dependence of $T_{\mathrm{BKT}}$ (c) and mean-field temperature $T_{\mathrm{%
MF}}$ (d) on $h_{z}$. The black dashed line denotes the exceptional point.
The line colors in (d) have the same notation as in (c). $h_{x}=0.1E_{F}$ in
all figures.}
\label{fig3}
\end{figure}

From Eq.~(\ref{eq5}), the intersection points between $J$ and the black
dashed line $y=2T/\pi $ in Fig.~\ref{fig2}(b) correspond to the BKT
transition points. Clearly, $T_{\mathrm{BKT}}$ is larger in the $\mathcal{PT}
$-broken region. In Fig.~\ref{fig3}(a), the dependence of $T_{\mathrm{BKT}}$
on the two-body binding energy $E_{B}$ at different $h_{z}$ is shown. The $%
T_{\mathrm{BKT}}$ in the $\mathcal{PT}$-broken region is always larger than
that in the $\mathcal{PT}$-symmetric region for small binding energy and
converges to $E_{F}/8$ in the strong coupling limit that is independent of
the Zeeman fields. Surprisingly, in the $\mathcal{PT}$-broken region, $T_{%
\mathrm{BKT}}$ for weak coupling can be much larger than $E_{F}/8$, which
was not discovered previously in Hermitian systems~\cite{SSBotelho,
Bertaina, MGong2,YXu,JPADevreese}. In both weak and strong coupling regions,
the order parameter only changes slightly (Fig.~\ref{fig3} (b)). The unusual
enlargement of $T_{\mathrm{BKT}}$ is determined by the non-Hermitian Zeeman
field $h_{z}$, which enhances the fermionic superfluidity not through the
order parameter.

In the $\mathcal{PT}$-broken region, $T_{\mathrm{BKT}}$ also oscillates with
$h_{z}$ around the equilibrium value $T_{0}=E_{F}/8$ for a fixed $E_{B}$, as
shown in Fig.~\ref{fig3}(c). The oscillation amplitude is reduced for a
large $E_{B}$. The oscillation period $\delta h^{\prime }=2\pi \times
E_{F}/8=\frac{\pi }{4}E_{F}$ is independent of $E_{B}$.
In Fig.~\ref{fig3}(d), we plot the mean-field temperature $T_{\mathrm{MF}}$
with respect to $h_{z}$, which increases as $h_{z}$ or $E_{B}$ increases. $%
T_{\mathrm{MF}}$ in the $\mathcal{PT}$-broken region is also larger than
that in the symmetric region. However, $T_{\mathrm{MF}}$ does not show a
periodic dependence on $h_{z}$ in the $\mathcal{PT}$-broken region. Another
critical temperature is the vortex lattice melting temperature $T_{\mathrm{%
vortex}}$. $T_{\mathrm{vortex}}$ in the $\mathcal{PT}$-broken region is
larger than that in $\mathcal{PT}$-symmetric region, and shows an
oscillation with $h_{z}$ but with a very small amplitude~\cite{SM}.

\begin{figure}[tb]
\centering
\includegraphics[width=0.5\textwidth]{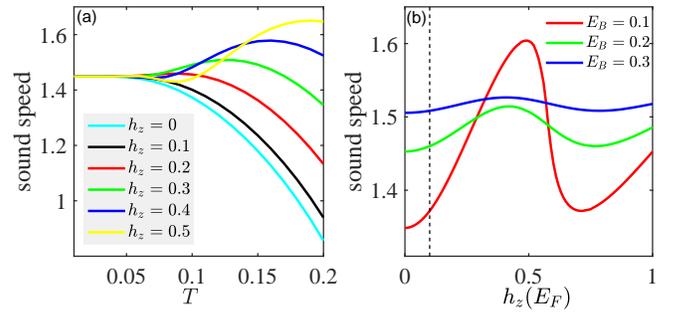}\newline
\caption{Dependence of the sound speed on temperature $T$ at different $h_{z}
$ in (a) and on $h_{z}$ at $T_{\mathrm{BKT}}$ in (b). $E_{B}=0.1E_{F}$ in
(a) and $h_{x}=0.1E_{F}$ in both figures.}
\label{fig4}
\end{figure}

In Fig. \ref{fig4}, we plot the sound speed $c=\sqrt{J/P}$ with respect to $T
$ and $h_{z}$. Similar to the superfluid density, at high temperature, the
sound speed in the $\mathcal{PT}$-broken region is always larger than that
in the $\mathcal{PT}$-symmetric region. For a fixed binding energy $E_{B}$,
the sound speed at $T_{\mathrm{BKT}}$ oscillates with $h_{z}$ with the
period same as that for $T_{\mathrm{BKT}}$ (Fig. \ref{fig3}(c)); while at a
fixed temperature, the sound speed oscillates with $h_{z}$ with the period
same as that for superfluid density (Fig. \ref{fig2}(a)). The measurement of
enhanced sound speed and its oscillation in the $\mathcal{PT}$-broken region
can be used to detect the symmetry-breaking physics~\cite%
{Christodoulou,TOzawa,LASidorenkov,LSalasnich,MBohlen}.

\emph{Discussion and summary}: In the experimental realization of the $%
\mathcal{PT}$ symmetry Hamiltonian, we can denote two atomic hyperfine
states (e.g., $^{2}S_{1/2},F=1/2$ ground-state manifold of $^{6}$Li) as the
effective spin states. The real in-plane Zeeman field $h_{x}$ can be
realized using radio-frequency wave or microwave coupling. The imaginary
Zeeman field can be achieved by introducing imbalanced loss to the two
spins, which can be done by resonantly coupling the spin-down (or spin-up)
state to the excited state using an optical beam. Usually, this process
leads to an imaginary Zeeman field possessing $\mathcal{PT}$-symmetry (given
by the loss rate difference between two spins)~\cite{JLi}, together with
some additional global loss term $-\frac{i\gamma _{g}}{2}\sigma _{0}$ ($%
\sigma _{0}$ $2\times 2$ is spin identity matrix).
Fortunately, this term can be gauged out by a unitary transformation on
inverse Green function $G^{-1}$ with the operator $U^{\prime
}=e^{iM\gamma_{g} \tau /2}$~\cite{SM}. Effectively, this transformation
replaces the order parameter $\phi $ with $\phi ^{\prime }=\phi
e^{-i\gamma_{g}\tau }$ in Eq.~(\ref{eq2}), which does not affect the action
in Eq.~(\ref{eq3}) and superfluid parameters~\cite{SM}. If we treat $\tau $
as the imaginary time, we see that $\phi ^{\prime }$ will decay at a rate $%
\gamma _{g}$ as expected, indicating that we are discussing the
instantaneous physics in a dissipative frame. This global loss would also
lead to total atomic density decay, leading to the decay in energy and
momentum units $E_{F}=n\pi /m$ and $k_{F}$ (which means that the binding
energy and Zeeman fields effectively increase with time). However, the
enhancement of the instantaneous BKT transition temperature persists.
Moreover, in realistic experiments, it is possible to couple the system with
a reservoir (which can supply atoms or BCS pairs) to compensate the global
atom loss instantaneously and thus maintain the atom density at some fixed
value (\textit{i.e.}, reaching equilibrium), yielding fixed BKT transition
temperature.

In summary, we investigate the BKT phase transition in a non-Hermitian $%
\mathcal{PT}$ symmetric 2D degenerate Fermi gas subject to an imaginary
Zeeman field and find an enhanced BKT superfluidity with a very large BKT
transition temperature at the weak coupling regime, in sharp contrast to the
Hermitian system. Although we focus on the BKT physics with spin balanced
atom density, it is possible to generalize the result to spin imbalanced
\cite{Mitra, Ong} or spin-orbit coupled 2D Fermi gases \cite{YJLin,PWang},
where non-Hermitian FFLO and/or topological states \cite{CZhang,WZhang,CQu}
may exist. Generalization to other BKT systems, such as Bose gases and 2D XY
model could be another interesting direction. Our work showcases a
surprising interplay between low-dimensional topological defects and
non-Hermitian effects, paving the way for studying non-Hermitian
low-dimensional phase transitions.

\begin{acknowledgments}
\textbf{Acknowledgements:} This work is supported by AFOSR (Grant No.
FA9550-20-1-0220), NSF (Grant No. PHY-1806227), and ARO (Grant No.
W911NF-17-1-0128).
\end{acknowledgments}

\newpage

\begin{widetext}
\setcounter{figure}{0} \renewcommand{\thefigure}{S\arabic{figure}} %
\setcounter{equation}{0} \renewcommand{\theequation}{S\arabic{equation}}

\section{Supplementary Materials}

\subsection{S1. Real physical quantities in PT-broken phase}


From the main text, the effective action at the saddle point after
integrating out the Fermi fields is given by
\begin{equation}
S_{\mathrm{eff}}=\beta (\frac{|\phi |^{2}}{U}+\sum_{\mathbf{p}}\xi _{\mathbf{%
p}})-\frac{1}{2}\mathrm{Tr}[\ln G^{-1}],
\end{equation}%
with the inverse Green function $G^{-1}=-\partial _{\tau }-H_{B}$. Replacing
$\partial _{\tau }$ with $-i\omega _{n}$, we get
\begin{equation}
S_{\mathrm{eff}}=\beta (\frac{|\phi |^{2}}{U}+\sum_{\mathbf{p}}\xi _{\mathbf{%
p}})-\frac{1}{2}\sum_{n,\mathbf{p}}\ln \{\prod_{\pm }[(i\omega _{n}-E_{\pm
})(i\omega _{n}+E_{\pm })]\},
\end{equation}%
where $\omega _{n}=\frac{(2n+1)\pi }{\beta }$ and $E_{\pm }=\sqrt{\xi _{%
\mathbf{p}}^{2}+|\phi |^{2}}\pm \sqrt{h_{x}^{2}-h_{z}^{2}}$. The eigenvalues
$E_{\pm }$ always locate at the right half-plane in the complex plane. After
summation over the Matsubara frequency,
\begin{equation}
S_{\mathrm{eff}}=\beta (\frac{|\phi |^{2}}{U}+\sum_{\mathbf{p}}\xi _{\mathbf{%
p}})-\sum_{\mathbf{p}}\{\ln [1+e^{-2\beta E_{\mathbf{p}}}+2e^{-\beta E_{%
\mathbf{p}}}\cosh (\beta h_{\mathrm{eff}})]+\beta E_{\mathbf{p}}\}
\end{equation}%
$E_{\mathbf{p}}=\sqrt{\xi _{\mathbf{p}}^{2}+|\phi |^{2}}$, $h_{\mathrm{eff}}=%
\sqrt{h_{x}^{2}-h_{z}^{2}}$. In the $\mathcal{PT}$-symmetric region, $%
h_{x}>h_{z}$, the effective action is real. In the $\mathcal{PT}$-broken
region, $h_{x}<h_{z}$, $h_{\mathrm{eff}}=ih^{\prime }$ with $h^{\prime }=%
\sqrt{h_{z}^{2}-h_{x}^{2}}$, and the effective action becomes
\begin{equation}
S_{\mathrm{eff}}=\beta (\frac{|\phi |^{2}}{U}+\sum_{\mathbf{p}}\xi _{\mathbf{%
p}})-\sum_{\mathbf{p}}\{\ln [1+e^{-2\beta E_{\mathbf{p}}}+2e^{-\beta E_{%
\mathbf{p}}}\cos (\beta h^{\prime })]+\beta E_{\mathbf{p}}\}.
\end{equation}%
The term inside the logarithmic function is always larger than $0$, and the
effective action at the saddle point in the $\mathcal{PT}$-broken region is
also real. As a result, the partition function is real and positive.

Now we prove the superfluid parameters $J$, $P$, $Q$ are real in the $%
\mathcal{PT}$-broken region. They are given by
\begin{eqnarray}
J &=&\frac{1}{4m}\sum_{\mathbf{p}}[n_{\mathbf{p}}-\frac{\beta \mathbf{p}^{2}%
}{8m}\sum_{i=\pm }\mathrm{sech}^{2}\frac{\beta E_{i}}{2}],\qquad \\
P &=&\frac{1}{8}\sum_{\mathbf{p}}[\frac{\Delta ^{2}}{E_{\mathbf{p}}^{3}}%
\sum_{i=\pm }\tanh \frac{\beta E_{i}}{2}+\frac{\beta \xi _{\mathbf{p}}^{2}}{%
2E_{\mathbf{p}}^{2}}\sum_{i=\pm }\mathrm{sech}^{2}\frac{\beta E_{i}}{2}],
\end{eqnarray}%
\begin{equation}
Q=\sum_{\mathbf{p}}n_{\mathbf{p}},\qquad n_{\mathbf{p}}=1-\frac{\xi _{%
\mathbf{p}}}{2E_{\mathbf{p}}}\sum_{i=\pm }\tanh \frac{\beta E_{i}}{2}
\end{equation}%
From the expressions, we only need prove two terms $\sum_{i=\pm }\tanh \frac{%
\beta E_{i}}{2}$ and $\sum_{i=\pm }\mathrm{sech}^{2}\frac{\beta E_{i}}{2}$
are real. For the first term
\begin{eqnarray}
\sum_{i=\pm }\tanh \frac{\beta E_{i}}{2} &=&\tanh \frac{\beta (E_{+}+E_{-})}{%
2}(1+\tanh \frac{\beta E_{+}}{2}\tanh \frac{\beta E_{-}}{2})  \notag \\
&=&\tanh (\beta E_{\mathbf{p}})(1+\frac{\sinh \frac{\beta E_{+}}{2}\sinh
\frac{\beta E_{-}}{2}}{\cosh \frac{\beta E_{+}}{2}\cosh \frac{\beta E_{-}}{2}%
})  \notag \\
&=&\tanh (\beta E_{\mathbf{p}})[1+\frac{\cosh (\beta E_{\mathbf{p}})-\cosh
(\beta h_{\mathrm{eff}})}{\cosh (\beta E_{\mathbf{p}})+\cosh (\beta h_{%
\mathrm{eff}})}].
\end{eqnarray}%
In the $\mathcal{PT}$-broken region, $\cosh (\beta h_{\mathrm{eff}})=\cos
(\beta h^{\prime })$, and we obtain
\begin{equation}
\sum_{i=\pm }\tanh \frac{\beta E_{i}}{2}=\frac{2\sinh (\beta E_{\mathbf{p}})%
}{\cosh (\beta E_{\mathbf{p}})+\cos (\beta h^{\prime })},
\end{equation}%
which is real. For the second term,
\begin{eqnarray}
\sum_{i=\pm }\mathrm{sech}^{2}\frac{\beta E_{i}}{2} &=&\frac{2}{1+\cosh
(\beta E_{+})}+\frac{2}{1+\cosh (\beta E_{-})}  \notag \\
&=&\frac{4+2\cosh (\beta E_{+})+2\cosh (\beta E_{-})}{1+\cosh (\beta
E_{+})+\cosh (\beta E_{-})+\cosh (\beta E_{+})\cosh (\beta E_{-})}  \notag \\
&=&\frac{4+4\cosh (\beta E_{\mathbf{p}})\cosh (\beta h_{\mathrm{eff}})}{%
1+2\cosh (\beta E_{\mathbf{p}})\cosh (\beta h_{\mathrm{eff}})+[\cosh (2\beta
E_{\mathbf{p}})+\cosh (2\beta h_{\mathrm{eff}})]/2}.
\end{eqnarray}%
In the $\mathcal{PT}$-broken region, we obtain
\begin{equation}
\sum_{i=\pm }\mathrm{sech}^{2}\frac{\beta E_{i}}{2}=\frac{4\left[ 1+\cosh
(\beta E_{\mathbf{p}})\cos (\beta h^{\prime })\right] }{1+2\cosh (\beta E_{%
\mathbf{p}})\cos (\beta h^{\prime })+[\cosh (2\beta E_{\mathbf{p}})+\cos
(2\beta h^{\prime })]/2},
\end{equation}%
which is also real.

\subsection{S2. Lattice melting temperature $T_{\mathrm{vortex}}$}

In Fig.~S1, we plot the vortex lattice melting temperature $T_{\mathrm{v}}$
with respect to the binding energy $E_{B}$ and imaginary Zeeman field $h_{z}$%
. Similar as $T_{\mathrm{BKT}}$, $T_{\mathrm{vortex}}$ in the $\mathcal{PT}$%
-broken region is larger at the weak coupling regime and then converges to
the value $\frac{3E_{F}}{40\pi }\approx 0.0239E_{F}$ at the strong coupling
limit. $T_{\mathrm{vortex}}$ in the $\mathcal{PT}$-broken region can be
larger than the strong coupling limit value. In (b), we see $T_{\mathrm{%
vortex}}$ also shows an oscillating behavior for a given $E_{B}$ with the
equilibrium center $\frac{3E_{F}}{40\pi }$ in the $\mathcal{PT}$-broken
region. The oscillation period is $\delta h^{\prime }=2\pi \times \frac{%
3E_{F}}{40\pi }=\frac{3E_{F}}{20}$ and the oscillation amplitude of $T_{%
\mathrm{vortex}}$ is quite small.

\begin{figure}[h]
\centering
\includegraphics[width=0.75\textwidth]{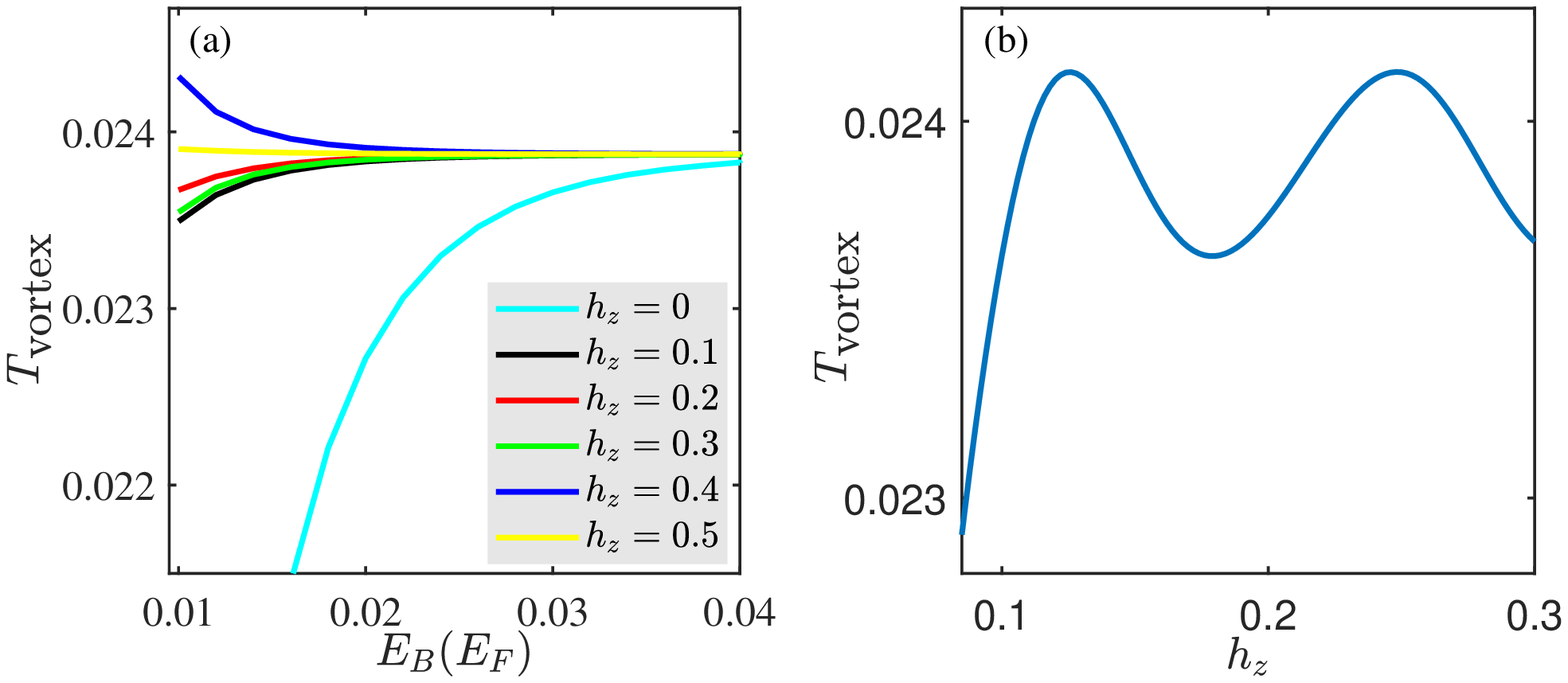}\newline
\caption{The vortex melting temperature versus $E_{B}$ (a) and $h_{z}$ (b),
respectively. In (b), $E_{B}=0.012E_{F}$. In (a) and (b), $h_{x}=0.1E_{F}$.}
\label{figs1}
\end{figure}

\section{S4. BKT physics with imbalanced gain and loss}

An exact $\mathcal{PT}$-symmetric Hamiltonian is usually hard to realize.
For example, a global loss term $-i\gamma \sigma _{0}$ on both spin species
will break the $\mathcal{PT}$ symmetry in the single-particle Hamiltonian $%
H_{s}(\mathbf{p})=\xi _{\mathbf{p}}+h_{x}\sigma _{x}+ih_{z}\sigma
_{z}-i\gamma \sigma _{0}$. Then the partition function $Z=\int D\bar{\phi}%
D\phi e^{-S_{\mathrm{eff}}[\bar{\phi},\phi ]}$ with the effective action $S_{%
\mathrm{eff}}=\int_{0}^{\beta }d\tau \int d\mathbf{r}(\frac{|\phi |^{2}}{U}%
+\sum_{\mathbf{p}}(\xi _{\mathbf{p}}-i\gamma ))-\frac{1}{2}\mathrm{Tr}[\ln
G^{-1}]$. The inverse Green function $G^{-1}=-\partial _{\tau }-H_{\mathrm{%
BdG}}$, where the BdG Hamiltonian in the Nambu basis is given by
\begin{equation}
H_{\mathrm{BdG}}=\left(
\begin{array}{cc}
H_{s}(\mathbf{p}) & -i\phi \sigma _{y} \\
i\phi ^{\ast }\sigma _{y} & -H_{s}(-\mathbf{p})%
\end{array}%
\right) .
\end{equation}

We can introduce a transformation $U^{-1}G^{-1}U$ with $U=\exp (iM\gamma
\tau )$ and $M=diag(1,1,-1,-1)$ to gauge out the diagonal $\gamma $ terms.
The effective action becomes $S_{\mathrm{eff}}=\int_{0}^{\beta }d\tau \int d%
\mathbf{r}(\frac{|\phi |^{2}}{U}+\sum_{\mathbf{p}}\xi _{\mathbf{p}})-\frac{1%
}{2}\mathrm{Tr}[\ln (G^{-1})^{\prime }]$ with $(G^{-1})^{\prime }=-\partial
_{\tau }-H_{\mathrm{BdG}}^{\prime }$ and
\begin{equation}
H_{\mathrm{BdG}}^{\prime }=\left(
\begin{array}{cccc}
\xi _{\mathbf{p}}+ih_{z} & h_{x} & 0 & -\phi ^{\prime } \\
h_{x} & \xi _{\mathbf{p}}-ih_{z} & \phi ^{\prime } & 0 \\
0 & \phi ^{\prime \ast } & -\xi _{\mathbf{p}}-ih_{z} & -h_{x} \\
-\phi ^{\prime \ast } & 0 & -h_{x} & -\xi _{\mathbf{p}}+ih_{z}%
\end{array}%
\right) .  \label{eq2}
\end{equation}%
with $\phi ^{\prime }=\phi e^{-2i\gamma \tau }$. The Nambu basis is now $%
\Psi =(e^{-i\gamma \tau }\Psi _{\mathbf{p}\uparrow },e^{-i\gamma \tau }\Psi
_{\mathbf{p}\downarrow },e^{i\gamma \tau }\Psi _{-\mathbf{p}\downarrow
}^{\dagger },e^{i\gamma \tau }\Psi _{-\mathbf{p}\uparrow }^{\dagger })^{T}$.
This means we can construct a $\mathcal{PT}$-symmetric system in the
dissipative frame since $\tau $ is the imaginary time. In this
transformation, $\phi \phi ^{\ast }=\phi ^{\prime }\phi ^{\prime \ast }$. As
a result, the effective action in Eq.~(3) and the superfluid parameters in
the main text will stay the same. Notice that the $\mathcal{PT}$-symmetric
non-Hermitian term $ih_{z}\sigma _{z}$ cannot be gauged out by the above
method.

\end{widetext}

\end{document}